\documentclass[conference,transmag]{IEEEtran}

\ifCLASSINFOpdf
   \usepackage[pdftex]{graphicx}
  
\else
 
\fi
\usepackage{amsmath}

\usepackage{algorithm}
\usepackage{algpseudocode}
\usepackage{lipsum}
\usepackage{float}
\usepackage{cite}
\usepackage{caption}
\usepackage{subcaption}
\usepackage{comment}
\usepackage{adjustbox}

\usepackage{xcolor}
\usepackage{acronym}

\usepackage{fixltx2e}

\acrodef{ADC}[ADC]{Analog to Digital Converter}
\acrodef{ADEXP}[AdExp-I\&F]{Adaptive-Exponential Integrate and Fire}
\acrodef{AER}[AER]{Address-Event Representation}
\acrodef{AEX}[AEX]{AER EXtension board}
\acrodef{AE}[AE]{Address-Event}
\acrodef{AFM}[AFM]{Atomic Force Microscope}
\acrodef{AGC}[AGC]{Automatic Gain Control}
\acrodef{AI}[AI]{Artificial Intelligence}
\acrodef{ALD}[ALD]{Atomic Layer Deposition}
\acrodef{AMDA}[AMDA]{AER Motherboard with D/A converters}
\acrodef{ANN}[ANN]{Artificial Neural Network}
\acrodef{API}[API]{Application Programming Interface}
\acrodef{ARM}[ARM]{Advanced RISC Machine}
\acrodef{ASIC}[ASIC]{Application Specific Integrated Circuit}
\acrodef{AdExp}[AdExp-IF]{Adaptive Exponential Integrate-and-Fire}
\acrodef{BCI}[BCI]{Brain-Computer-Interface}
\acrodef{BCM}[BCM]{Bienenstock-Cooper-Munro}
\acrodef{BD}[BD]{Bundled Data}
\acrodef{BEOL}[BEOL]{Back-end of Line}
\acrodef{BG}[BG]{Bias Generator}
\acrodef{BMI}[BMI]{Brain-Machince Interface}
\acrodef{BTB}[BTB]{band-to-band tunnelling}
\acrodef{BP}[BP]{Back-propagation}
\acrodef{BPTT}[BPTT]{Back-propagation Through Time}
\acrodef{CAD}[CAD]{Computer Aided Design}
\acrodef{CAM}[CAM]{Content Addressable Memory}
\acrodef{CAVIAR}[CAVIAR]{Convolution AER Vision Architecture for Real-Time}
\acrodef{CA}[CA]{Cortical Automaton}
\acrodef{CCN}[CCN]{Cooperative and Competitive Network}
\acrodef{CDR}[CDR]{Clock-Data Recovery}
\acrodef{CFC}[CFC]{Current to Frequency Converter}
\acrodef{CHP}[CHP]{Communicating Hardware Processes}
\acrodef{CMIM}[CMIM]{Metal-insulator-metal Capacitor}
\acrodef{CML}[CML]{Current Mode Logic}
\acrodef{CMP}[CMP]{Chemical Mechanical Polishing}
\acrodef{CMOL}[CMOL]{Hybrid CMOS nanoelectronic circuits}
\acrodef{CMOS}[CMOS]{Complementary Metal-Oxide-Semiconductor}
\acrodef{CNN}[CCN]{Convolutional Neural Network}
\acrodef{COTS}[COTS]{Commercial Off-The-Shelf}
\acrodef{CPG}[CPG]{Central Pattern Generator}
\acrodef{CPLD}[CPLD]{Complex Programmable Logic Device}
\acrodef{CPU}[CPU]{Central Processing Unit}
\acrodef{CSM}[CSM]{Cortical State Machine}
\acrodef{CSP}[CSP]{Constraint Satisfaction Problem}
\acrodef{CV}[CV]{Coefficient of Variation}
\acrodef{DAC}[DAC]{Digital to Analog Converter}
\acrodef{DAS}[DAS]{Dynamic Auditory Sensor}
\acrodef{DAVIS}[DAVIS]{Dynamic and Active Pixel Vision Sensor}
\acrodef{DBN}[DBN]{Deep Belief Network}
\acrodef{DBS}[DBS]{Deep-Brain Stimulation}
\acrodef{DFA}[DFA]{Deterministic Finite Automaton}
\acrodef{DIBL}[DIBL]{drain-induced-barrier-lowering}
\acrodef{DI}[DI]{delay insensitive}
\acrodef{DMA}[DMA]{Direct Memory Access}
\acrodef{DNF}[DNF]{Dynamic Neural Field}
\acrodef{DNN}[DNN]{Deep Neural Network}
\acrodef{DoF}[DoF]{Degrees of Freedom}
\acrodef{DPE}[DPE]{Dynamic Parameter Estimation}
\acrodef{DPI}[DPI]{Differential Pair Integrator}
\acrodef{DRRZ}[DR-RZ]{Dual-Rail Return-to-Zero}
\acrodef{DRAM}[DRAM]{Dynamic Random Access Memory}
\acrodef{DR}[DR]{Dual Rail}
\acrodef{DSP}[DSP]{Digital Signal Processor}
\acrodef{DVS}[DVS]{Dynamic Vision Sensor}
\acrodef{DYNAP}[DYNAP]{Dynamic Neuromorphic Asynchronous Processor}
\acrodef{EBL}[EBL]{Electron Beam Lithography}
\acrodef{ECoG}[ECoG]{Electrocorticography}
\acrodef{EDVAC}[EDVAC]{Electronic Discrete Variable Automatic Computer}
\acrodef{EEG}[EEG]{electroencephalography}
\acrodef{EIN}[EIN]{Excitatory-Inhibitory Network}
\acrodef{EM}[EM]{Expectation Maximization}
\acrodef{EPSC}[EPSC]{Excitatory Post-Synaptic Current}
\acrodef{EPSP}[EPSP]{Excitatory Post-Synaptic Potential}
\acrodef{ET}[ET]{Eligibility Trace}
\acrodef{EZ}[EZ]{Epileptogenic Zone}
\acrodef{FDSOI}[FDSOI]{Fully-Depleted Silicon on Insulator}
\acrodef{FEOL}[FEOL]{Front-end of Line}
\acrodef{FET}[FET]{Field-Effect Transistor}
\acrodef{FFT}[FFT]{Fast Fourier Transform}
\acrodef{FI}[F-I]{Frequency-Current}
\acrodef{FPGA}[FPGA]{Field Programmable Gate Array}
\acrodef{FR}[FR]{Fast Ripple}
\acrodef{FSA}[FSA]{Finite State Automaton}
\acrodef{FSM}[FSM]{Finite State Machine}
\acrodef{GIDL}[GIDL]{gate-induced-drain-leakage}
\acrodef{GOPS}[GOPS]{Giga-Operations per Second}
\acrodef{GPU}[GPU]{Graphical Processing Unit}
\acrodef{GUI}[GUI]{Graphical User Interface}
\acrodef{HAL}[HAL]{Hardware Abstraction Layer}
\acrodef{HFO}[HFO]{High Frequency Oscillation}
\acrodef{HH}[H\&H]{Hodgkin \& Huxley}
\acrodef{HMM}[HMM]{Hidden Markov Model}
\acrodef{HRS}[HRS]{High-Resistive State}
\acrodef{HR}[HR]{Human Readable}
\acrodef{HSE}[HSE]{Handshaking Expansion}
\acrodef{HW}[HW]{Hardware}
\acrodef{IBCI}[IBCI]{Implantable BCI}
\acrodef{ICT}[ICT]{Information and Communication Technology}
\acrodef{IC}[IC]{Integrated Circuit}
\acrodef{ICL}[ICL]{Implantable Closed Loop}
\acrodef{IEEG}[iEEG]{intracranial electroencephalography}
\acrodef{IF2DWTA}[IF2DWTA]{Integrate \& Fire 2--Dimensional WTA}
\acrodef{IFSLWTA}[IFSLWTA]{Integrate \& Fire Stop Learning WTA}
\acrodef{IF}[I\&F]{Integrate-and-Fire}
\acrodef{IMU}[IMU]{Inertial Measurement Unit}
\acrodef{INCF}[INCF]{International Neuroinformatics Coordinating Facility}
\acrodef{INI}[INI]{Institute of Neuroinformatics}
\acrodef{INRC}[Intel NRC]{Intel Neuromorphic Research Community}
\acrodef{IO}[I/O]{Input/Output}
\acrodef{IoT}[IoT]{Internet of Things}
\acrodef{IPSC}[IPSC]{Inhibitory Post-Synaptic Current}
\acrodef{IPSP}[IPSP]{Inhibitory Post-Synaptic Potential}
\acrodef{IP}[IP]{Intellectual Property}
\acrodef{ISI}[ISI]{Inter-Spike Interval}
\acrodef{IoT}[IoT]{Internet of Things}
\acrodef{JFLAP}[JFLAP]{Java - Formal Languages and Automata Package}
\acrodef{LEDR}[LEDR]{Level-Encoded Dual-Rail}
\acrodef{LFP}[LFP]{Local Field Potential}
\acrodef{LFSR}[LFSR]{Linear Feedback Shift Register}
\acrodef{LIF}[LIF]{Leaky Integrate and Fire}
\acrodef{LLC}[LLC]{Low Leakage Cell}
\acrodef{LNA}[LNA]{Low-Noise Amplifier}
\acrodef{LPF}[LPF]{Low Pass Filter}
\acrodef{LRS}[LRS]{Low-Resistive State}
\acrodef{LSM}[LSM]{Liquid State Machine}
\acrodef{LTD}[LTD]{Long Term Depression}
\acrodef{LTI}[LTI]{Linear Time-Invariant}
\acrodef{LTP}[LTP]{Long Term Potentiation}
\acrodef{LTU}[LTU]{Linear Threshold Unit}
\acrodef{LUT}[LUT]{Look-Up Table}
\acrodef{LVDS}[LVDS]{Low Voltage Differential Signaling}
\acrodef{MD}[MD]{Medical Device}
\acrodef{MCMC}[MCMC]{Markov-Chain Monte Carlo}
\acrodef{MEMS}[MEMS]{Micro Electro Mechanical System}
\acrodef{MFR}[MFR]{Mean Firing Rate}
\acrodef{MIM}[MIM]{Metal Insulator Metal}
\acrodef{ML}[ML]{Machine Leanring}
\acrodef{MLP}[MLP]{Multilayer Perceptron}
\acrodef{MOSCAP}[MOSCAP]{Metal Oxide Semiconductor Capacitor}
\acrodef{MOSFET}[MOSFET]{Metal Oxide Semiconductor Field-Effect Transistor}
\acrodef{MOS}[MOS]{Metal Oxide Semiconductor}
\acrodef{MRI}[MRI]{Magnetic Resonance Imaging}
\acrodef{NDFSM}[NDFSM]{Non-deterministic Finite State Machine} 
\acrodef{ND}[ND]{Noise-Driven}
\acrodef{NEF}[NEF]{Neural Engineering Framework}
\acrodef{NHML}[NHML]{Neuromorphic Hardware Mark-up Language}
\acrodef{NIL}[NIL]{Nano-Imprint Lithography}
\acrodef{NLP}[NLP]{Natural Language Processing}
\acrodef{NMDA}[NMDA]{N-Methyl-D-Aspartate}
\acrodef{NME}[NE]{Neuromorphic Engineering}
\acrodef{NN}[NN]{Neural Network}
\acrodef{NRZ}[NRZ]{Non-Return-to-Zero}
\acrodef{NSM}[NSM]{Neural State Machine}
\acrodef{OR}[OR]{Operating Room}
\acrodef{OTA}[OTA]{Operational Transconductance Amplifier}
\acrodef{PCB}[PCB]{Printed Circuit Board}
\acrodef{PCHB}[PCHB]{Pre-Charge Half-Buffer}
\acrodef{PCM}[PCM]{Phase Change Memory}
\acrodef{PD}[PD]{Parkinson Disease}
\acrodef{PE}[PE]{Phase Encoding}
\acrodef{PFA}[PFA]{Probabilistic Finite Automaton}
\acrodef{PFC}[PFC]{prefrontal cortex}
\acrodef{PFM}[PFM]{Pulse Frequency Modulation}
\acrodef{PM}[PM]{Personalized Medicine}
\acrodef{PR}[PR]{Production Rule}
\acrodef{PSC}[PSC]{Post-Synaptic Current}
\acrodef{PSP}[PSP]{Post-Synaptic Potential}
\acrodef{PSTH}[PSTH]{Peri-Stimulus Time Histogram}
\acrodef{PVD}[PVD]{Physical Vapor Deposition }
\acrodef{QDI}[QDI]{Quasi Delay Insensitive}
\acrodef{RAM}[RAM]{Random Access Memory}
\acrodef{RDF}[RDF]{random dopant fluctuation}
\acrodef{RELU}[ReLu]{Rectified Linear Unit}
\acrodef{RLS}[RLS]{Recursive Least-Squares}
\acrodef{RMSE}[RMSE]{Root Mean Squared-Error}
\acrodef{RMS}[RMS]{Root Mean Squared}
\acrodef{RNN}[RNN]{Recurrent Neural Network}
\acrodef{ROLLS}[ROLLS]{Reconfigurable On-Line Learning Spiking}
\acrodef{RRAM}[R-RAM]{Resistive Random Access Memory}
\acrodef{R}[R]{Ripples}
\acrodef{SAC}[SAC]{Selective Attention Chip}
\acrodef{SAT}[SAT]{Boolean Satisfiability Problem}
\acrodef{SCX}[SCX]{Silicon CorteX}
\acrodef{SD}[SD]{Signal-Driven}
\acrodef{SDSP}[SDSP]{Spike Driven Synaptic Plasticity}
\acrodef{SEM}[SEM]{Spike-based Expectation Maximization}
\acrodef{SLAM}[SLAM]{Simultaneous Localization and Mapping}
\acrodef{SNN}[SNN]{Spiking Neural Network}
\acrodef{SNR}[SNR]{Signal to Noise Ratio}
\acrodef{SOC}[SOC]{System-On-Chip}
\acrodef{SOI}[SOI]{Silicon on Insulator}
\acrodef{SoA}[SoA]{state-of-the-art}
\acrodef{SP}[SP]{Separation Property}
\acrodef{SRAM}[SRAM]{Static Random Access Memory}
\acrodef{STDP}[STDP]{Spike-Timing Dependent Plasticity}
\acrodef{STD}[STD]{Short-Term Depression}
\acrodef{STEM}[STEM]{Science, Technology, Engineering and Mathematics}
\acrodef{STP}[STP]{Short-Term Plasticity}
\acrodef{STT-MRAM}[STT-MRAM]{Spin-Transfer Torque Magnetic Random Access Memory}
\acrodef{STT}[STT]{Spin-Transfer Torque}
\acrodef{SW}[SW]{Software}
\acrodef{TCAM}[TCAM]{Ternary Content-Addressable Memory}
\acrodef{TFT}[TFT]{Thin Film Transistor}
\acrodef{TPU}[TPU]{Tensor Processing Unit}
\acrodef{TRL}[TRL]{Technology Readiness Level}
\acrodef{USB}[USB]{Universal Serial Bus}
\acrodef{VHDL}[VHDL]{VHSIC Hardware Description Language}
\acrodef{VLSI}[VLSI]{Very Large Scale Integration}
\acrodef{VOR}[VOR]{Vestibulo-Ocular Reflex}
\acrodef{WCST}[WCST]{Wisconsin Card Sorting Test}
\acrodef{WTA}[WTA]{Winner-Take-All}
\acrodef{XML}[XML]{eXtensible Mark-up Language}
\acrodef{CTXCTL}[CTXCTL]{CortexControl}
\acrodef{divmod3}[DIVMOD3]{divisibility of a number by three}
\acrodef{hWTA}[hWTA]{hard Winner-Take-All}
\acrodef{sWTA}[sWTA]{soft Winner-Take-All}
\acrodef{APMOM}[APMOM]{Alternate Polarity Metal On Metal}

\title{ALIVE}

\begin{document}

\IEEEdisplaynontitleabstractindextext
\newcommand{\MP}[1]{{\color{red}  Melika: #1}}
\newcommand{\AR}[1]{{\color{green}  Arianna: #1}}
\newcommand{\MC}[1]{{\color{blue}  Matteo: #1}}
\newcommand{\GI}[1]{{\color{cyan} Giacomo: #1}}

\title{Stochastic dendrites enable online learning in mixed-signal neuromorphic processing systems}

\author{\IEEEauthorblockN{
Matteo Cartiglia\IEEEauthorrefmark{1}\IEEEauthorrefmark{3},
Arianna Rubino\IEEEauthorrefmark{1}\IEEEauthorrefmark{3}, 
Shyam Narayanan\IEEEauthorrefmark{1},  
Charlotte Frenkel \IEEEauthorrefmark{1}, 
Germain Haessig \IEEEauthorrefmark{2}, \\ 
Giacomo Indiveri \IEEEauthorrefmark{1}, 
Melika Payvand \IEEEauthorrefmark{1}}

\IEEEauthorblockA{
\IEEEauthorrefmark{1}Institute of Neuroinformatics,
University of Zurich and ETH Zurich, Winterthurerstr. 190, 8057 Zurich, Switzerland  \\
\IEEEauthorrefmark{2}Austrian Institute of Technology
Giefinggasse 4, 1210 Vienna, Austria
\IEEEauthorrefmark{3} These authors have contributed equally to this work. \\
Email: \{rubinoa, camatteo, melika\}@ini.uzh.ch }
}


\maketitle
\begin{abstract}
  The stringent memory and power constraints required in edge-computing sensory-processing applications have made event-driven neuromorphic systems a promising technology.
  On-chip online learning  provides such systems the ability to learn the statistics of the incoming data and to adapt to their changes.
  Implementing online learning on event driven-neuromorphic systems requires
  (i) a spike-based learning algorithm that calculates the weight updates using only local information from streaming data,
  (ii) mapping these weight updates onto limited bit precision memory
  and (iii) doing so in a robust manner that does not lead to unnecessary updates as the system is reaching its optimal output. Recent neuroscience studies have shown how dendritic compartments of cortical neurons can solve these problems in biological neural networks. 
  Inspired by these studies we propose spike-based learning circuits to implement stochastic dendritic online learning.
  The circuits are embedded in a prototype spiking neural network fabricated using a 180\,nm process.
  Following an algorithm-circuits co-design approach we present circuits and behavioral simulation results that demonstrate the learning rule features.
  We validate the proposed method using behavioral simulations of a single-layer network with 4-bit precision weights applied to the MNIST benchmark, and demonstrating results that reach accuracy levels above 85\%. 
\end{abstract}


\section{Introduction}

Our society is shifting to an era of pervasive specialized ``edge computing" for a wide variety of tasks. 
Artificial Intelligence (AI) is fueling this trend by achieving remarkable results in pattern recognition.
However, the conventional computing technology used to run AI algorithms is not ideal for edge-computing tasks that require minimal power consumption and always-on real-time processing.
Inspired by the computational principles of neural processing systems, neuromorphic technologies have the potential to satisfy the power-efficient and real-time processing requirements of edge-computing applications~\cite{Indiveri_Liu15}. 
The power efficiency of neuromorphic computing systems is enabled by co-locating memory and processing~\cite{Sebastian_etal20} and by implementing fine-grain parallelism strategies that remove the need to time-multiplex data processing. This is achieved by instantiating multiple copies of the processing elements (i.e., synapses and neurons) which have temporal dynamics that are well-matched to those of the sensory signals of interest and which operate in continuous physical time~\cite{Rubino_etal20,Indiveri_Sandamirskaya19}.
Following this approach several mixed-signal neuromorphic processing systems have been proposed that implement spiking neural networks~\cite{Thakur_etal18}. However, endowing these systems with online learning abilities remains an open challenge.

\begin{figure}
 \vspace*{-0.3cm}

\centering
\includegraphics[scale=0.4]{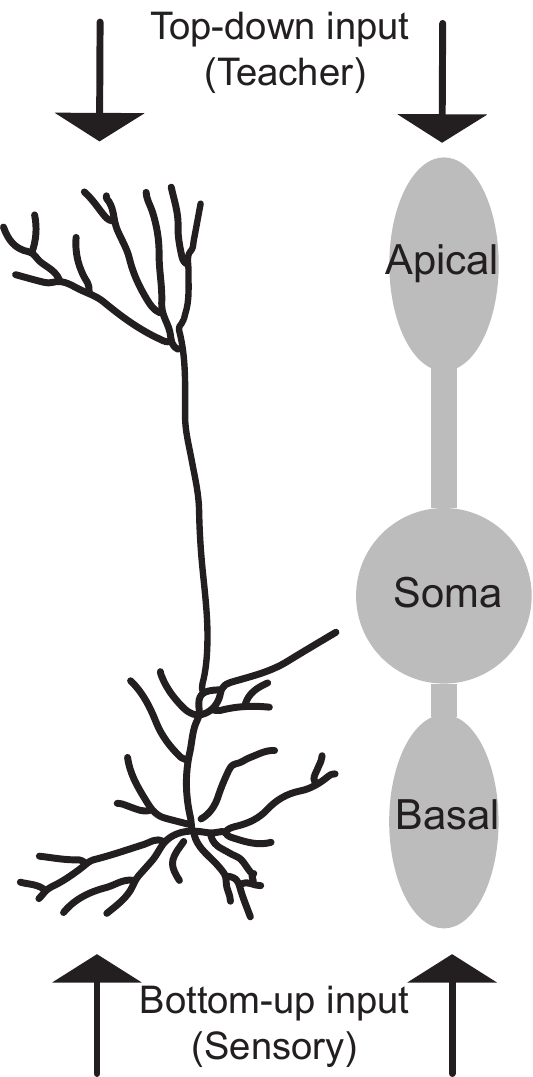}
\caption{Depiction of a pyramidal cortical neuron (left) and its multi-compartment model used in artificial neural networks (right). Adapted from~\cite{guerguiev_etal2017}.}
\label{fig:dendrites}
 \vspace*{-0.3cm}

\end{figure}

Since providing a top-down error signal has been very successful in deep learning\cite{Crafton_etal19,Zenke_Neftci21}, some neuromorphic implementations have recently focused on using errors for online learning~\cite{Payvand_etal20, Cartiglia_etal20}. Moreover, many neuroscientific studies have recently focused on error-based learning in the brain~\cite{Sacramento_etal18, whittington_boga2017}. 
In these studies, an error at the output units is assessed by comparing the self-generated activity of the neurons with a target activity. This error is then sent to the neurons as a feedback signal. Neurons are modeled with multiple compartments with several active dendritic branches, each directly linked to a somatic compartment (Fig.~\ref{fig:dendrites}). The feedback signals arrive on the distal Apical dendrites, while proximal Basal dendrites receive feed-forward sensory information. Such types of models have been shown to implement error-backpropagation using biologically-plausible plasticity mechanisms~\cite{Sacramento_etal18,whittington_boga2017,guerguiev_etal2017}.

\begin{figure*}
\centering
\includegraphics[width=\textwidth]{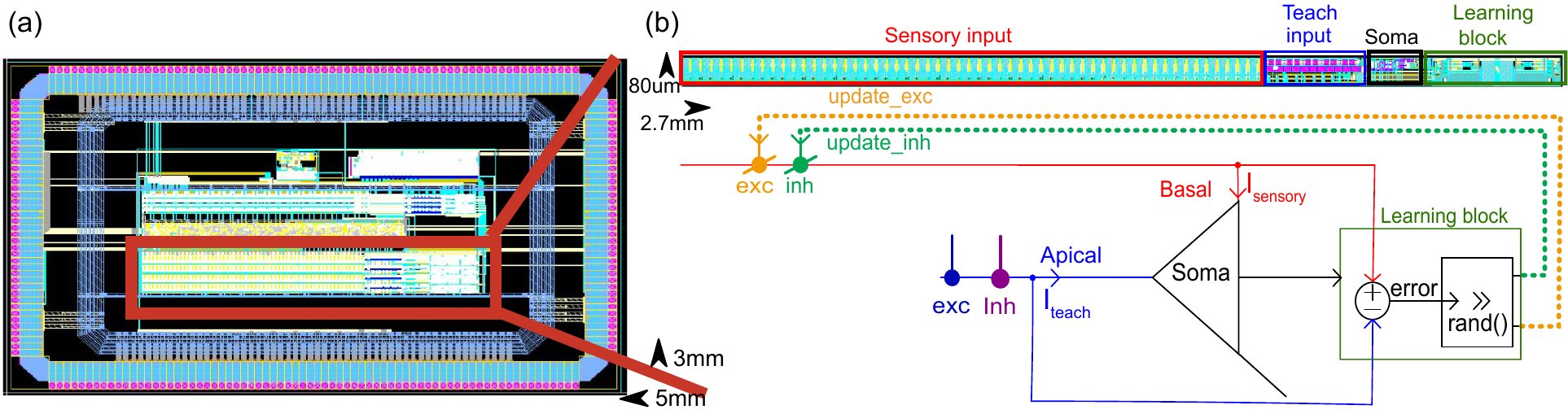}
\caption{ (a) A layout-view of the 3x5 $mm^2$ design. The circuits proposed are highlighted in the image and are composed of 4 individual neurons. (b) A close-up of a single neuron. The dendritic branches, the soma, and the learning block are all highlighted both in the schematic view and in the physical layout. The Sensory input is made up of 40 excitatory  and 16 inhibitory plastic synapses. The Teach input is made up of 3 excitatory and 5 inhibitory non-plastic synapses.}
\label{fig:neuron+chip}
\end{figure*}

Inspired by these studies, we developed mixed-signal electronic circuits that implement power-efficient online plasticity mechanisms that can be interfaced to neuromorphic synapse and neuron circuits. Specifically, since the synaptic weight precision is limited by the area of the memory that can be implemented on-chip, an interface circuitry is required to translate the analog weight updates to these low-bit precision memories. To solve this, we implement stochastic rounding on chip, which converts the analog weight updates to a probability of synaptic weight change ~\cite{Muller_Indiveri15,Courbariaux_etal15}. Additionally, this mechanism provides robustness to the weight update process by reducing the probability of updates as the error decreases to near-zero values.

We describe the software-hardware co-design approach that validated the circuit designs with behavioral simulations, and that led to the fabrication of the circuits using a standard 180\,nm \ac{CMOS} process (Fig.~\ref{fig:neuron+chip}).
In the next sections we present the learning algorithm, the circuits implementations, and finally, the results from the hardware and software simulations.
\vspace{-0.1cm}

\section{The algorithm}
\vspace{-0.1cm}

\subsection{The multi-compartment neuron model}

The selected neuron model used is a multi-compartment \ac{LIF} neuron in which currents from two dendritic branches, the bottom-up Sensory branch ($I_{Sensory}$) and the top-down Teacher branch ($I_{Teach}$), get integrated into a somatic compartment ($I_{Soma}$). This is summarized in Eq.~\ref{eq::eq_soma}. 

\begin{equation}
\tau_{Soma} \cfrac{ d I_{Soma}}{d t} + I_{Soma}(t)=  {{I_{Sensory} (t)} + {I_{Teach}(t)}}
\label{eq::eq_soma}
\end{equation}

where $\tau_{Soma}$ is the neuron's time constant. 

The Sensory and Teach dendritric dynamics are modeled by Eq. ~\ref{eq::eq_dend} and \ref{eq::teach} respectively. 

\begin{equation}
\tau_{Sensory}\cfrac{ d {I_{Sensory_j}}}{d t} + {I_{Sensory_j}} = \sum\limits_i \sum \limits_n {I_{w_{ij}}} \delta_{i,s} (t - t_n)
\label{eq::eq_dend}
\end{equation}

\begin{equation}
\tau_{Teach}\cfrac{ d {I_{Teach_j}}}{d t} + {I_{Teach_j}} =  \sum \limits_n  {I_{w_{Teach}}} \delta_{i,t} ( t - t_n)
\label{eq::teach}
\end{equation}

where $i,j$ represents the input and output neurons respectively, and $n$ is the time at which the input neurons spike. The $\delta_{i,s}$ and $\delta_{i,t}$ function describes the input spike trains at the Sensory and Teach branches, respectively. 
$\tau_{Sensory}$ and $\tau_{Teach}$ represent the respective Sensory and Teach dendritic time constants, $I_{w_{ij}}$ is the plastic synaptic strength between Sensory input $i$ and output neuron $j$ (Basal input), and ${I_{w_{Teach}}}$ is the non-plastic constant teacher current (Apical input).

\vspace{-0.1cm}

\subsection{The learning rule} 
\label{LR_section}

The learning rule follows a dendritic-based prediction learning~\cite{Urbanczik_Senn14} which resembles Delta rule. The Delta rule is the simplest form of gradient-descent learning, minimizing a single-layer neural network's Least Mean Square (LMS) error. 
Here, the error is defined as the difference between the Sensory and Teach dendritic branches as is described in Eq.~\ref{eq::learning rule}.

\begin{equation}
  { err_{ij}} = \alpha({I_{Teach_j}} - {I_{Sensory_j}})x_{i}|I_{Teach_j}|
  \label{eq::learning rule}
\end{equation}

where $\alpha$ is the learning rate and the error is calculated every time a pre-synaptic spike occurs ($x_i$). This error is scaled by the absolute value of$(I_{Teach_j}$, which is in case of a true targets, is positive and negative otherwise. In this way if the teaching signal is not present (i.e., $I_{Teach_j}=0$) the learning will stop.

To manage the changes on the limited-bit precision synaptic memory (here 4 bits) with the analog weight updates, that are proportional to the error signal, we employed a stochastic rounding mechanism~\cite{Muller_Indiveri15, Courbariaux_etal15, Frenkel_etal20}. 
This is implemented by comparing the magnitude of the error to a random number (here 6 bits), and if greater, incrementing or decrementing the weight depending on the sign of the error. Otherwise, no weights are updated. 
Therefore, the probability of weight update is proportional to the error in the network.
This is summarized in Eq.~\ref{eq::rand}.


\begin{equation}
\Delta w_{ij} (@x_i) = 
\begin{cases}
  \pm 1\quad \text{if}\quad |err_{ij}| > rand()\\
  0\quad \text{otherwise}
\label{eq::rand}
\end{cases}
\end{equation}

\begin{figure}
 \vspace*{-0.3cm}

\centering
\includegraphics[scale=0.7]{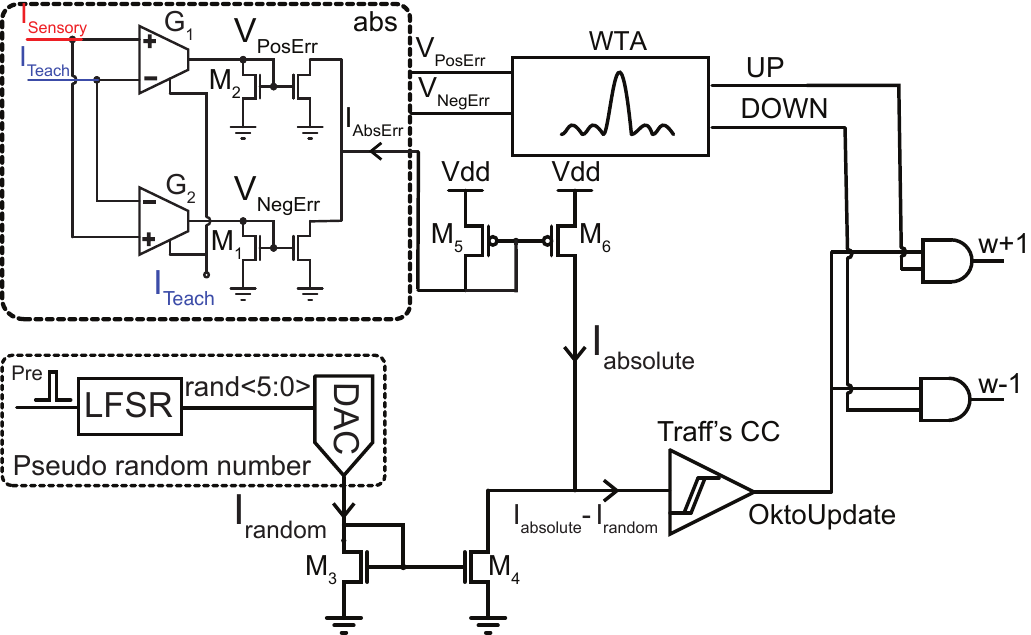}
\caption{Block diagram of the online learning mechanism. It consists of error generation and stochastic rounding.}
\label{fig:circuit_diagram}
 \vspace*{-0.3cm}

\end{figure}

\section{The neuromorphic system}

\subsection{The network architecture}

The circuits have been fabricated in a prototype chip using a 180\,nm \ac{CMOS} technology. The network occupies an area of $15 mm^2$ ($5\,mm$ x $3 mm$). The total area occupied by the neural core presented here is $1.01 mm^2$. The physical layout of the network is shown in Fig.~\ref{fig:neuron+chip}a.\\
The neural core can be broken down into four identical rows, where each row has one neuron and 64 plastic and non-plastic synapses (Fig.~\ref{fig:neuron+chip}b). 
The total synaptic activity consists of two main dendritic branches. 
The Sensory branch (Basal, ${I_{Sensory}}$) has 56 independent excitatory and inhibitory plastic synapses (4 bit resolution), while the Teach branch (Apical, ${I_{Teach}}$), has 8 excitatory and inhibitory non-plastic synapses. The excitatory synapses inject a positive current into the dendrite, while the inhibitory ones takes current away from it. Therefore, the total current in both dendritic branches is the subtraction between the excitatory and inhibitory currents.\\

Upon the arrival of any input event on the Sensory dendritic branch, the Soma integrates the weighted input, and if it reaches a threshold, it generates an output spike. Moreover, the Sensory input event activates the learning block which computes the error, generates a random number, and sends update signals if the former is greater than the latter.

\subsection{The learning circuits}

Figure~\ref{fig:circuit_diagram} illustrates the block diagram of the learning mechanism, which consists of the error-generation and stochastic rounding blocks.

\subsubsection{Error generation} \label{error_gen}

The error generation implements Eq.~\ref{eq::learning rule}, getting ${I_{Teach}}$ and ${I_{Sensory}}$ as inputs. The magnitude and the direction of weight update are determined by the absolute value ($abs$), and Winner-Takes-All (WTA) circuit, respectively.

The $abs$ circuit calculates the error by subtracting ${I_{Teach}}$ and ${I_{Sensory}}$ from each other in two directions, using $G_1$ and $G_2$ transconductance amplifiers, giving rise to the positive (${V_{PosErr}}$) and negative (${V_{NegErr}}$) errors.
Depending on which input is greater, either the $M_1$ or the $M_2$ transistor is on, whose current flows in $I_{AbsErr}$ giving rise to the absolute value of the error~\cite{Payvand_etal19}. 
The $abs$ circuit is biased with ${I_{Teach}}$ to gate the generation of the error signals only when the teacher is presented. If the teacher is not present (${I_{Teach}}$ = 0), the $abs$ circuit is OFF. The Teach input can be either positive or negative, however the $abs$ circuit is biased always with the unsigned ${I_{Teach}}$. The learning block is designed to distinguish between a positive and a negative Teach input and change the synaptic weights accordingly.

The WTA decides the direction of update ($UP$ or $DOWN$) depending on which one of ${V_{PosErr}}$ and ${V_{NegErr}}$ is higher~\cite{Payvand_Indiveri19}. 
If ${V_{PosErr}}$ and ${V_{NegErr}}$ are equal, $UP$ and $DOWN$ output of the WTA circuit are both set to low, and thus no update signal is generated. This makes the learning more robust by avoiding unnecessary updates as the system is reaching its optimal output.

\subsubsection{Stochastic rounding circuits}

The stochastic rounding block implements Eq.~\ref{eq::rand}.
Every time a pre-synaptic spike is generated, a  6-bit pseudo-random number is generated from an on-chip \ac{LFSR} and is converted into an analog current (${I_{random}}$) using a 6-bit current \ac{DAC}. 
The  ${I_{random}}$ current is then compared to the error current, $I_{absolute}$, generated by the error generation block (see \ref{error_gen}). 
The comparison happens through a modified version of Traff's current comparator (CC) proposed in \cite{Traff_1992}. 
Traff's comparator input voltage is determined by the difference between ${I_{absolute}}$ and ${I_{random}}$ and is initially around ${Vdd/2}$.  
As soon as ${I_{absolute}}$ and ${I_{random}}$ are not equal, the input voltage increases or decreases and the change is amplified by the CC.
The output ${OktoUpdate}$ is set to high/low if $I_{absolute}$ is larger/less than $I_{random}$. The ${OktoUpdate}$ signal allows the weight update signals ($UP$ and $DOWN$) to pass through, only when it is high. 

Since the probability of generating ${OktoUpdate}$ is proportional to the error, as the error is reduced, the probability of generating an update is also reduced. This leads to less weight updates as the system is reaching the optimal output which makes the learning more robust and power efficient.

It is worth noting that the event-driven nature of the input allows us to use a single stochastic rounding block for the entire row and generate an update signal on the arrival of any pre-synaptic sensory input events.

\section{Results}

\subsection{Circuit simulations}
\begin{figure}[!]
\vspace*{-0.3cm}

  \begin{subfigure}{0.25\textwidth}
    \centering
    \includegraphics[width=\linewidth]{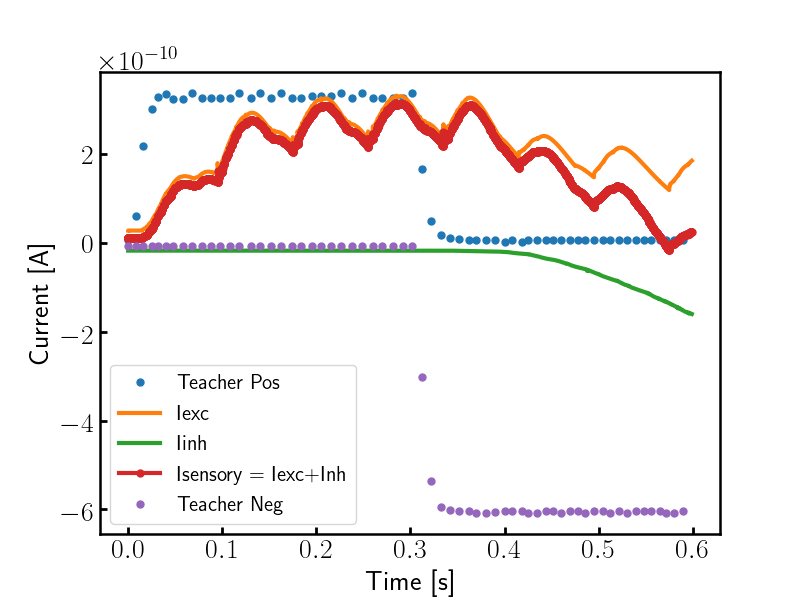}
    \subcaption{}

  \end{subfigure}
  \vspace*{-0.2cm}
  \begin{subfigure}{0.25\textwidth}
    \centering
    \includegraphics[width=\linewidth]{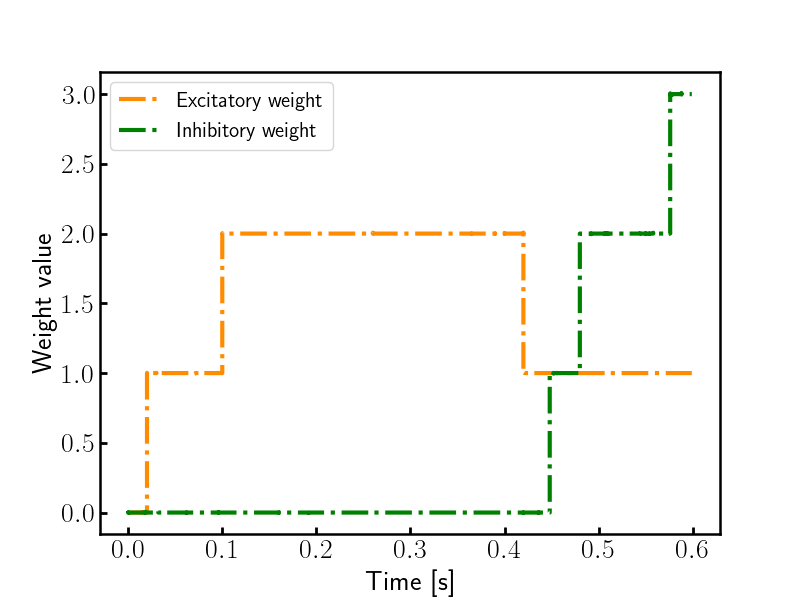}
    \subcaption{}

  \end{subfigure}

      \caption{Cadence simulations of the learning signals of a single neuron; (a) example of the signals involved in the learning process.  The sensory dendritic current (in red) is the sum of the excitatory (in orange) and inhibitory currents (in green). A positive teacher signal (dotted in blue) is presented until t = 0.3s. A negative teacher signal (dotted in purple) is presented after that.
 (b) weight changes associated with the learning of (a). The weights change their value to minimize the error between the sensory current and the teacher current. An excitatory (dashdotted in orange) and an inhibitory (dashdotted in green) weight are shown. }
 \vspace*{-0.3cm}

    \label{fig:traces}

\end{figure}

To validate the learning rule at a schematic-level, we show the simulation results from one of the four identical rows of the chip (Figure~\ref{fig:traces}).
Initially, a positive teacher (dotted blue in Fig.~\ref{fig:traces}a) is presented, which causes the weights of the excitatory synapses (dashdotted in orange in Fig.~\ref{fig:traces}a) to strengthen and minimize the error between the teacher and the sensory branch (shown in red in Fig.~\ref{fig:traces}a). At t = 0.3s, a negative teacher is shown to the network, which causes the strengthened excitatory synapses to weaken, and the inhibitory synapses (dashdotted in green in Fig.~\ref{fig:traces}b) to strengthen and minimize the error between the two dendritic branches.

Table~\ref{tab::specs} shows a direct comparison with previous works using on-chip on-line error-based learning~\cite{Frenkel_etal20,Davies_etal18}. 
The area in ~\cite{Frenkel_etal20} has been normalized to a 180\,nm technology node to allow for a better comparison with our work.
The energy per update in ~\cite{Davies_etal18} is not directly comparable to our work, since the weight update is calculated in a microcontroller, and the reported figure does not take that into account. The 720 pJ reported in this work, is the total energy in one row consumed to calculate and update one bit of synaptic memory. 

\begin{table}
\centering
\resizebox{\linewidth}{!}{%

\begin{tabular}{ cccc }

    		\textbf{Feature}& \textbf{\cite{Frenkel_etal20}}
    		& \textbf{\cite{Davies_etal18}}
    		&\textbf{This work}\\
		 \hline
            Resources  & C5×5@10 & Re-configurable & 4 neurons\\
            or topology &–FC128–FC10& &64 synapse\\
            \hline
            Implementation & Digital & Digital & Mixed-signal\\
            \hline
            Learning & on-line &  on-line & on-line \\& error-based & re-configurable & error-based
            \\
        \hline
		Synaptic\\ resolution & 8-bit& 1-9 bits & 4-bit\\
		\hline
		Static power & 61$\mu{W}$&not-applicable\footnotemark[1] &455\,nW\\ consumption &&& per neuron\\
		\hline
		Energy metric & N/A
		&120 \,pJ\footnotemark[1] 
		&  720\,pJ \\
		&
		& per weight update &per weight update\\
		\hline
		Technology node & 28\,nm FDSOI&14 \,nm FinFET& 180\,nm CMOS\\
		\hline
		Power supply& 0.6\,V & 0.75\,V& 1.8\,V \\
		\hline
		Area & $13.2\,mm^2$ \footnotemark[2] &not-applicable \footnotemark[1] &$15\,mm^2$ \\
		
        \hline

\end{tabular}}

\footnotesize[1]{The learning block is implemented using a micro controller: area and power figures are not directly comparable.}\\
\footnotesize[2]{Design normalized to a 180\,nm technology node.}

\caption{Comparison with previous works using on-chip on-line error-based learning.}
\label{tab::specs}
  \vspace*{-0.2cm}

\end{table}

\subsection{System}

\begin{figure}[!]

\vspace*{-0.3cm}

   \begin{subfigure}[b]{0.24\textwidth}
    \centering
    \includegraphics[width=\textwidth]{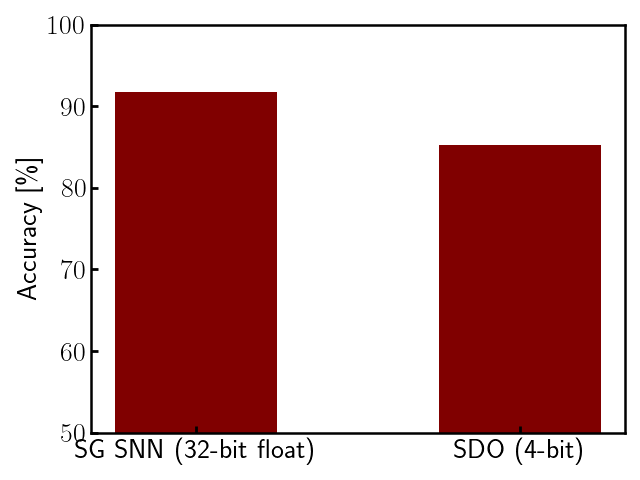}
    \subcaption{}
  \end{subfigure}
   \begin{subfigure}[b]{0.24\textwidth}
    \centering
  \includegraphics[width=\textwidth]{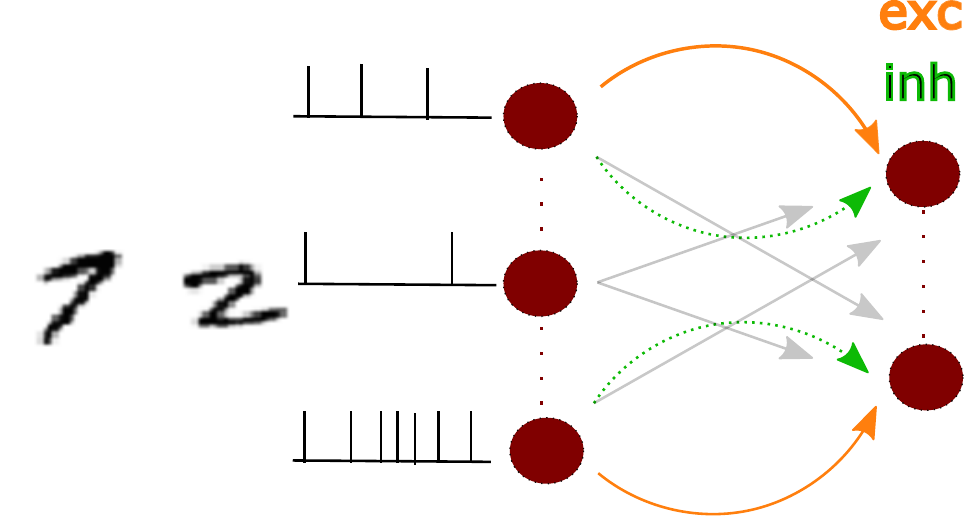}
    \subcaption{}
  \end{subfigure}
  \caption{MNIST benchmarking results. (a) Accuracy comparison of a 784-10 SNN network trained using surrogate gradients (SG) versus the same network trained using the proposed stochastic dendritic online (SDO)  rule. 80-20 randomized division was used between training and testing. (b) Topology of the network in the SDO simulations: input pixels are connected to the output layer with excitatory (in orange) and inhibitory (in green) connections. }
  \label{fig:benchmarks}
   \vspace*{-0.3cm}

\end{figure}

The proposed learning rule and system were fully modeled using the BRIAN2 simulator~\cite{Goodman_Brette08}. Fig.~\ref{fig:benchmarks} shows the results of a shallow network (with topology 784-10 neurons) on the MNIST handwritten digit dataset. It shows a direct comparison with a \ac{SNN} trained with surrogate gradient descent \cite{Neftci_etal19} using 32-bit floating point (SG SNN), while the proposed implementation uses 4-bit weights. The SG SNN reaches 91\% accuracy while the stochastic dendritic online implementation reaches above 85\%. Interestingly~\cite{Frenkel_etal19a} achieves similar results with the same network topology.
 
Although both are \acp{SNN}, the difference in performance can be attributed to that (i) the hardware network implements weight updates on streaming data, with a batch size of one, compared to a batch size of 256 in SG SNN, and (ii) the hardware uses only 4 bits of precision compared to the floating point resolution in the SG SNN case.
\section{Discussion and Conclusion}

We presented a prototype neuromorphic chip designed and fabricated in 180\,nm technology which implements stochastic dendritic-based online learning.  
We proposed an algorithm--circuits co-design approach and validated it with circuit simulations that demonstrate the proper operation of the learning rule. 

We benchmarked our system via system-level simulations, and obtained above 85\% accuracy on the MNIST dataset using only a one-layer network with 4-bit weight precision. 

As our circuits show that single neurons can adapt based on their incoming activity, they represent a valid candidate for adaptive ``edge computing'' applications that require online learning. To this end, future work will include interfacing the chip with signals that need to be classified at the edge, such as biomedical or industrial signals.


\section*{Acknowledgment}
The authors would like to thank Junren Chen for support during the last steps in the chip layout. This project has received funding from the European Union's H2020 research and innovation programme under the H2020 BeFerrosynaptic (871737), MeM-Scales project (871371) and Sinergia Project (CRSII5-180316).

\bibliographystyle{unsrt}

\newpage

\bibliography{biblio/biblioncs.bib, biblio/bib_extra.bib}

\end{document}